\documentclass[useAMS,galley]{mn2e}
\usepackage{graphicx}
\usepackage{times,mathptm}
\newcommand{\ltsima}{$\; \buildrel < \over \sim \;$}
\newcommand{\simlt}{\lower.5ex\hbox{\ltsima}}
\newcommand{\gtsima}{$\; \buildrel > \over \sim \;$}
\newcommand{\simgt}{\lower.5ex\hbox{\gtsima}}

\newcommand{\cgs}{ ${\rm erg~cm}^{-2}~{\rm s}^{-1}$} 
\newcommand{\lum}{\rm erg s$^{-1}$}

\def\lesssim{\mathrel{\hbox{\rlap{\hbox{\lower4pt\hbox{$\sim$}}}\hbox{$<$}}}}
\def\gtrsim{\mathrel{\hbox{\rlap{\hbox{\lower4pt\hbox{$\sim$}}}\hbox{$>$}}}}

\def\arcmin{\hbox{$^\prime$}}
\def\arcsec{\hbox{$^{\prime\prime}$}}

\def\ab1450{$AB_{1450(1+z)}$}

\def\xray{\hbox{X-ray}}

\def\oiii{\hbox{[O\ {\sc iii}}]}
\def\lsun{\hbox{$L_\odot$}}
\def\loiii{$L_{[O\ III]}$}
\newcommand\phn{\phantom{0}}%
\newcommand\phs{\phantom{$-$}}%

\def\chandra{{\it Chandra\/}}

\def\heao1{{\it HEAO-1\/}}

\def\spitzer{{\it Spitzer\/}}
\def\scuba{{\it SCUBA\/}}

\def\rosat{{\it ROSAT\/}}

\def\xmm{{XMM-{\it Newton\/}}}

\def\aj{AJ}
\def\araa{ARA\&A}
\def\apj{ApJ}
\def\apjl{ApJ}
\def\apjs{ApJS}

\def\aap{A\&A}

\def\mnras{MNRAS}

\title[THE X-RAY PROPERTIES OF TYPE~II QUASAR CANDIDATES FROM THE SDSS]
{Evidence for X-ray Obscuration in Type~II Quasar Candidates from the 
Sloan Digital Sky Survey}
\author[C. Vignali et al.]
{
Cristian Vignali,$^{1,2}$\thanks{E-mail: cristian.vignali@bo.astro.it (CV); dma@ast.cam.ac.uk (DMA); andrea.comastri@bo.astro.it (AC).}
Dave M. Alexander$^{3}$\footnotemark[1] and 
Andrea Comastri$^{1}$\footnotemark[1] \\ \\ 
$^{1}$ INAF -- Osservatorio Astronomico di Bologna, Via Ranzani 1, 
40127 Bologna, Italy \\
$^{2}$ Dipartimento di Astronomia, Universit\`a degli Studi di Bologna, 
Via Ranzani 1, 40127 Bologna, Italy \\
$^{3}$ Institute of Astronomy, Madingley Road, Cambridge, CB3~0HA \\
}

\begin{document}

\date{Accepted 2004 ???. Received 2004 ???; in original form 2004 ???}

\pagerange{\pageref{firstpage}--\pageref{lastpage}} \pubyear{2004}

\maketitle

\label{firstpage}

\begin{abstract}
Recently, Zakamska et al. (2003) selected 291 high-ionization narrow 
emission-line AGN in the redshift range 0.3--0.83 from the Sloan 
Digital Sky Survey spectroscopic data. 
The sample includes both Seyfert~II galaxies and their higher luminosity 
``cousins'', Type~II quasar candidates. 
Here we present the results on the \xray\ properties of 17 
of these objects for which archival \xray\ data (\rosat\ and \xmm) 
are available. 
Three sources have been significantly ($\simgt6\sigma$) detected, 
one being the most radio-loud source of the 
sample; its \xray\ emission, possibly enhanced by jet emission, is consistent 
with the absence of absorption. 
Another source has a $\approx6\sigma$ detection in the 
\rosat\ All-Sky Survey, possibly complex radio structure, 
and no evidence for strong \xray\ absorption. For the third \xray\ detection, 
the \xmm\ spectrum indicates a column density of 
\hbox{$N_{\rm H}=1.26^{+0.75}_{-0.51}\times10^{22}$~cm$^{-2}$}; this 
result, coupled with the \hbox{2--10~keV} luminosity of 
$\approx$~4$\times10^{44}$~\lum, makes this source a genuine Type~II quasar. 
Using the \oiii$\lambda$5007 line luminosities, we estimated the intrinsic 
\xray\ power of the AGN and found that $\gtrsim47$ per cent 
of the observed sample shows indications of \xray\ absorption, 
with column densities $\gtrsim10^{22}$~cm$^{-2}$. 
This provides further evidence that a considerable fraction are 
obscured quasars. Support to our conclusions also comes from the 
recent analysis of RASS data performed by Zakamska et al. (2004),
 who found five 
additional lower significance \hbox{($\approx2.1\sigma-3.6\sigma$)} 
\xray\ matches.  
\end{abstract}


\section{Introduction}
One of the basic predictions of Unification models of active galactic nuclei 
(AGN) is the existence of a population of luminous, highly obscured 
(i.e., with high-ionization, narrow optical emission-line spectra) 
AGN, the so-called Type~II quasars, often referred to as the higher 
luminosity counterparts of local Seyfert~II galaxies. 
In the \xray\ band these objects are expected to be highly luminous 
($\gtrsim$~a~few~$\times10^{44}$~\lum) and absorbed by 
column densities $\gtrsim10^{22}$~cm$^{-2}$; 
given their properties, they would constitute an important ingredient of most 
AGN synthesis models for the \xray\ background 
(XRB; e.g., Comastri et al. 2001; Gilli, Salvati \& Hasinger 2001).
Before the advent of the current generation of \xray\ telescopes 
(\chandra\ and \xmm), Type~II quasars formed quite an elusive 
class of sources [but see Brandt et al. (1997) and Franceschini et 
al. (2000)]. 
Recent \xray\ surveys have found several genuine Type~II 
quasars (e.g., Norman et al. 2002; Stern et al. 2002; 
Fiore et al. 2003; Perola et al. 2004; Gandhi et al. 2004; 
Caccianiga et al. 2004; Szokoly et al. 2004) 
and many more candidate Type~II quasars 
(e.g., Crawford et al. 2001, 2002; Mainieri et al. 2002; Barger et al. 2003). 
Multiwavelength observations of these objects generally support their 
Type~II quasar status; in particular, at optical wavelengths Type~II quasars 
often show high equivalent width, narrow emission lines 
with high-ionization line ratios. 
However, in some cases 
broad-band, high signal-to-noise ratio optical/near-infrared 
follow-up observations of \xray\ selected 
Type~II quasar candidates have revealed broad components in their permitted 
emission lines (e.g., Halpern et al. 1999; Akiyama et al. 2002). 
Overall, less than $\approx$~20 per cent 
of the hard \xray, presumably absorbed 
sources found in moderately deep/ultra-deep surveys have optical 
counterparts characterized by high-ionization narrow emission lines 
(e.g., Barger et al. 2003), suggesting that the Type~II quasar 
population is not adequately sampled at present. 
Furthermore, the optical counterparts of \xray\ selected Type~II quasar 
candidates are typically so faint that the spectroscopic identification is 
quite a challenging task. Therefore, to have a comprehensive view of 
the properties of Type~II quasars, the selection of a relatively optically 
bright sample is necessary. 
In this regard, recently Zakamska et al. (2003; hereafter Z03) selected 291 
high-ionization narrow emission line AGN in the redshift range 
\hbox{0.3--0.83} from the Sloan Digital Sky Survey (SDSS; York et al. 
2000) spectroscopic data. 
The sample includes both Seyfert~II galaxies and 
their higher luminosity ``cousins'', Type~II quasar candidates. 
This is the only published optically selected sample of candidate 
Type~II quasars in this redshift range.\footnote{ 
We note that Djorgovski et al. (2001) have produced a sample of obscured 
quasar candidates, selected on the basis of narrow permitted emission lines 
and high-ionization line ratios. This sample covers the 
complementary 0.31--0.36 redshift range.} 
The main goal of the present work is to 
define the basic \xray\ properties of optically selected Type~II 
quasar candidates. 

Hereafter we adopt the ``concordance'' (WMAP) cosmology 
($H_{0}$=70~km~s$^{-1}$~Mpc$^{-1}$, $\Omega_{\rm M}$=0.3, and 
$\Omega_{\Lambda}$=0.7; Spergel et al. 2003).

\section{Sample selection}
To select candidate obscured (i.e., Type~II) AGN from the 
SDSS spectroscopic data, Z03 searched for objects with narrow emission 
lines without underlying broad 
components and with line ratios characteristic of non-stellar ionizing 
radiation. The redshift range \hbox{(0.3--0.83)} is chosen to include the 
\oiii$\lambda$5007 line in all of the spectra and this allows a statistically 
relevant study of the emission-line properties of the sources.
The sample members are characterized by spectra with signal-to-noise ratios 
\hbox{$\ge7.5$}, rest-frame equivalent width of the \oiii\ line 
\hbox{$\ge4$~\AA}, and \hbox{FWHM\ (H$\beta)<2000$~km~s$^{-1}$}. 
The high-ionization selection criteria adopted by Z03 
(see $\S$~3.3 of Z03 and Kewley et al. 2001 for further references) 
are designed to properly reject star-forming galaxies and 
genuinely unobscured objects, 
such as narrow-line Seyfert~1 galaxies (NLS1s), which otherwise would match 
the FWHM criterion (e.g., Williams, Pogge \& Mathur 2002). 
Given the width of the fibers used in the SDSS spectroscopic runs 
(3\arcsec), a significant contribution to the overall emission can 
come from the AGN host galaxies. 
To compute reliable AGN emission-line parameters, Z03 subtracted 
the host galaxy contribution from each spectrum (using templates) 
before computing the line parameters. 
About 50 per cent of the objects selected by Z03 have \oiii\ line luminosities 
in the range 3$\times10^{8}-10^{10}$~$\lsun$, comparable to those of 
luminous ($-27<M_{\rm B}<-23$) quasars. This, coupled with other evidences, 
suggests that at least the objects in the luminous subsample 
are Type~II quasar candidates.\footnote{
A widely accepted definition of Type~II quasars does not exist at present. 
From the optical perspective, Type~II quasars are the luminous [typically 
\hbox{$M_{\rm B}<-23$} (e.g., Schmidt \& Green 1983), 
although this ``historical'' absolute magnitude limit 
has not a physically supported motivation and is based on a different 
cosmology than the one adopted here] analogs of Seyfert~II galaxies. 
In the X-rays, objects with \hbox{2--10~keV} luminosity 
\hbox{$\gtrsim3\times10^{44}$~\lum} and characterized by absorption 
\hbox{$\gtrsim10^{22}$~cm$^{-2}$} are usually called Type~II quasars. 
We note that {\it a priori} it is possible that the 
objects optically classified as Type~II quasars do not match the \xray\ 
definition and viceversa. Therefore, in the following 
we will refer to our objects as Type~II quasar candidates.}

It must be noted that Z03 sample, is not complete; 
$\approx$~28 per cent of the 
sources have been selected as ``targets'', $\approx$~42 per cent as 
``serendipitous'', $\approx$~19 per cent from the Deep Southern Equatorial 
Scan (DSES) plates (where the magnitude limit is $\approx$~1 mag fainter), 
and 11 per cent from the ``special plates'' (see Z03 for details). 
Nevertheless, the careful and accurate analyses performed by Z03 to select 
the final sample of 291 candidate obscured AGN allow reliable 
studies of optically selected Type~II quasar candidates.

\section{The X-ray data: reduction and analysis}
The Z03 catalog of candidate Type~II quasars was cross correlated with 
archival \rosat, \chandra, and \xmm\ observations. 
Our basic \xray\ source searching strategy is performed using \rosat\ 
because of its large field-of-view. 
On the contrary, \chandra\ and \xmm\ cover smaller regions of the sky but 
provide greater sensitivity and broader energy range (up to $\approx$~10~keV). 
Seventeen sources lie in archival \xray\ observations. 
The observation log and source \xray\ fluxes (or upper limits) are presented 
in Table~1. 
%
\begin{table*}
\centering
\begin{minipage}{170mm}
\caption{Zakamska et al. (2003) Type~II AGN in \rosat\ 
and \xmm\ archival data.}
\scriptsize
\begin{tabular}{lccccccccccc}
\hline
Src$^{a}$ & SDSS~J & $z$ & $N_{\rm H, gal}$$^{b}$ & $S_{\rm 1.4~GHz}$$^{c}$ & 
$\log$~\loiii & $\log L_{\rm 2-10~keV}$$^{d}$ & Instr.$^{e}$ & 
$F_{\rm 0.5-2~keV}$$^{f}$ & Exp.~Time & Off-axis & Obs.$^{g}$ \\
ID \# & & & (10$^{20}$~cm$^{-2}$) & (mJy) & (\lsun) & & & & (ks) & 
Angle ($\arcmin$) & ID \\
\hline
{\phn}34   & 021047.01$-$100152.9 & 0.540 & 2.17 & $<0.48$ &  {\phn}9.79 & 45.1$\pm{0.6}$ & P & $<2.00$ &      14.6 &      19.6  &              800114p     \\
{\phn}55   & 023359.93$+$004012.7 & 0.388 & 2.82 & $<0.42$ &  {\phn}8.17 & 43.5$\pm{0.6}$ & P & $<0.43$ &      28.2 & {\phn}6.8  &              800482p     \\
{\phn}59   & 024309.79$+$000640.3 & 0.414 & 3.56 & $<0.84$ &  {\phn}7.95 & 43.3$\pm{0.6}$ & P & $<3.15$ & {\phn}5.5 &      10.3  &  {\phn}{\phs}150021p$-$2 \\
{\phn}59$^{h}$
           &                      &       &      &         &             &                & H & $<0.92$ &      72.0 & {\phn}9.5  &  {\phn}{\phs}701352h$-$1 \\
{\phn}68   & 025558.00$-$005954.0 & 0.700 & 6.33 & $<0.42$ &  {\phn}8.51 & 43.9$\pm{0.6}$ & P & $<3.54$ & {\phn}3.0 &      15.4  &              190338p     \\
{\phn}70   & 025951.28$+$002301.0 & 0.505 & 7.14 & $<0.42$ &  {\phn}8.53 & 43.9$\pm{0.6}$ & P & $<2.87$ & {\phn}5.0 &      16.4  &              700393p     \\
     130   & 084234.94$+$362503.1 & 0.561 & 3.41 &   1.64  &       10.10 & 45.5$\pm{0.6}$ & H & $<2.53$ &      28.1 & {\phn}5.7  &              800854h     \\
     148   & 090933.51$+$425346.5 & 0.670 & 1.60 & 4009.5  &  {\phn}8.92 & 44.3$\pm{0.6}$ & P &  $39.5$ &      22.1 & {\phn}0.2  &  {\phn}{\phs}700329p$-$1 \\
      152  & 092014.11$+$453157.3 & 0.402 & 1.51 & $<2.13$ &  {\phn}9.04 & 44.4$\pm{0.6}$ & P & $<2.75$ &  {\phn}4.3 &      12.0  &              700539p     \\
      174  & 100854.43$+$461300.7 & 0.544 & 0.95 &   7.16  &  {\phn}8.32 & 43.7$\pm{0.6}$ & H & $<3.42$ &  {\phn}9.5 &      17.5  &              702430h     \\
      188  & 104505.39$+$561118.4 & 0.428 & 0.65 &   2.59  &  {\phn}9.08 & 44.4$\pm{0.6}$ & P & $<3.51$ &  {\phn}3.9 &      14.0  &              600058p     \\
      204  & 122656.48$+$013124.3 & 0.732 & 1.84 & $<2.22$ &  {\phn}9.66 & 45.0$\pm{0.6}$ & P & $2.63$  &       24.8 &      12.5  &  {\phn}{\phs}600242p$-$1 \\
204$^{i}$  &                      &       &      &         &             & 44.6--44.7     & X & $3.90$  &   7.8--9.8 & {\phn}6.0  &                0110990201 \\
      208  & 123453.10$+$640510.2 & 0.594 & 1.87 & $<0.45$ &  {\phn}8.77 & 44.1$\pm{0.6}$ & P & $<4.74$ &  {\phn}3.0 &      17.0  &              800263p     \\
      209  & 124736.07$+$023110.7 & 0.487 & 1.76 &   0.90  &  {\phn}8.59 & 43.9$\pm{0.6}$ & P & $<3.61$ &  {\phn}5.5 &      17.5  &              700020p     \\
      212  & 130740.56$-$021455.3 & 0.425 & 1.75 &   1.66  &  {\phn}8.92 & 44.3$\pm{0.6}$ & P & $<8.37$ &  {\phn}1.6 &      18.2  &              200531p     \\
      239  & 150117.96$+$545518.3 & 0.338 & 1.40 &  20.87$^{j}$  
							   &  {\phn}9.06 & 44.4$\pm{0.6}$ & R &  $30.3$ &  {\phn}1.1 &            &              930725p     \\
      256  & 164131.73$+$385840.9 & 0.596 & 1.16 &   2.80  &  {\phn}9.92 & 45.3$\pm{0.6}$ & P & $<2.44$ &  {\phn}5.5 &      16.1  &              201538p     \\
      258  & 165627.28$+$351401.7 & 0.679 & 1.73 &   1.16  &  {\phn}8.57 & 43.9$\pm{0.6}$ & P & $<1.53$ &       22.1 &      18.1  &              400374p     \\
\hline
\end{tabular}
\end{minipage}
\begin{minipage}{170mm}
$^{a}$ From Table~1 of Zakamska et al. 2003. 
$^{b}$ From Dickey \& Lockman 1990. 
$^{c}$ Integrated 1.4~GHz flux density (or 3$\sigma$ upper limit) 
from FIRST (Becker, White \& Helfand 1995). 
$^{d}$ Predicted 2--10~keV luminosity and relative uncertainty 
(using the 1$\sigma$ scatter in the Mulchaey et al. 1994 correlation; 
see $\S$4 for details) in units of \lum. 
$^{e}$ ``P'' and ``H'' indicate the \rosat\ PSPC and HRI observations, 
respectively, while ``R'' indicates the \rosat\ All-Sky Survey (RASS) 
observation (in this case the off-axis angle is not reported); 
``X'' is referred to the object with \xmm\ observation. 
$^{f}$ Galactic absorption-corrected flux (or 3$\sigma$ upper 
limit) in the observed \hbox{0.5--2~keV} band, in units of 10$^{-14}$~\cgs. 
$^{g}$ Observation ID in the \rosat\ (\rosat\ observation request, ROR) 
and \xmm\ archives. 
$^{h}$ Both PSPC and HRI data are reported for source \#59, since 
the results obtained from their analyses appear significantly different 
(see text and Table~2 for details). 
$^{i}$ \xmm\ observation of source \#204. 
The flux and observed (de-absorbed) 2--10~keV luminosity shown here have been 
derived directly from \xray\ spectral fitting of \xmm\ data (see $\S$4.1); 
the range is due to slight pn--MOS differences in relative normalisations. 
Both the pn and MOS exposure times 
after the removal of periods of flaring background are reported. 
$^{j}$ The flux density is referred to the FIRST source which is closest 
(0.3\arcsec) to the optical position of the quasar. 
This object has possibly a peculiar radio morphology, as reported in the 
Appendix~A.4 and Fig.~2 of Zakamska et al. (2004). 
\end{minipage}
\end{table*}

%

\subsection{ROSAT observations}
At first the Z03 sample was cross correlated with \rosat\ data 
[pointed PSPC, pointed HRI, and the \rosat\ All-Sky Survey (RASS)]. 
We used only the central regions (20\arcmin\ radius) of the \rosat\ PSPC and 
HRI detectors, where the sensitivity is highest and the PSPC window support 
structure does not affect source detection. 
This maximizes the probability of detecting faint sources. 
In cases of multiple observations of the same SDSS quasar with the same \xray\ 
instrument, the one with the best combination of off-axis angle and 
exposure time has been chosen.
Typically, the use of PSPC data is preferred to that of HRI data, given the 
larger number of objects with PSPC data available.\footnote{For 
source \#59 we report the results from both the PSPC and HRI observations.} 
The detailed source detection procedure is described in 
Vignali, Brandt \& Schneider (2003). 
Briefly, source detection was performed in the \hbox{0.5--2~keV} band 
for the PSPC and in the full band \hbox{($\approx$~0.1--2.4~keV)} 
for the HRI.
A conservative matching radius of 40\arcsec\ was 
used to take into account the broadening of the Point Spread Function (PSF) 
at large off-axis angles,\footnote{See 
ftp://ftp.xray.mpe.mpg.de/rosat/catalogues/1rxp/wga\_rosatsrc.html.}
although typically a good match is obtained within 15--20\arcsec\ 
(e.g., Vignali et al. 2003). 

Standard likelihood detection algorithms available under {\sc MIDAS/EXSAS} 
(Zimmermann et al. 1998) have been used. 
We set the detection threshold to a likelihood 
of \hbox{$L=-\ln(P_{\rm e})=6$}, corresponding to a probability 
$P_{\rm e}$ of the order of 2.5$\times$10$^{-3}$ that the observed number 
of photons in the source cell is produced entirely by a background fluctuation 
(corresponding to the $\approx$~3.2$\sigma$ detection level; 
Cruddace, Hasinger \& Schmitt 1988). 

Sixteen of the Z03 SDSS AGN fall in the inner part of the 
field-of-view of \rosat\ pointed observations. 
Two have \xray\ detections 
(one in the PSPC and the other in both the PSPC and HRI); 
these \xray\ counterparts lie at 0\farcs1 (source \#148) 
and 8\farcs2 (source \#204) from the optical position; 
given their relatively large fluxes (see Table~1), 
we do not expect either of these sources to be spurious. 
We determined \xray\ count rates (for the two detections) and 3$\sigma$ upper 
limits for the 14 \xray\ non-detections (12 from PSPC and 2 from HRI) 
using the {\sc SOSTA} task in the {\sc XIMAGE} package 
(version 4.1; Giommi et al. 1992). For the two \xray\ 
detections the \xray\ count rates obtained using {\sc ximage} 
have been compared with those derived using the 
maximum likelihood method, finding good agreement. 
The \xray\ count rates have been converted using {\sc PIMMS} (Mukai 2001) 
into observed \hbox{0.5--2~keV} fluxes (for both PSPC and HRI; this energy 
range corresponds to \hbox{$\approx$~0.7--3.5~keV} in the source rest frame) 
using a power law with $\Gamma=2.0$ and Galactic absorption 
(from Dickey \& Lockman 1990). 
Although both Type~I (i.e., with broad permitted emission lines) 
and Type~II (i.e., with narrow permitted and forbidden emission lines) 
AGN in the local Universe are generally well described by 
\hbox{$\Gamma$=1.8--2.0} \xray\ continua 
(e.g., Nandra \& Pounds 1994; Reeves \& Turner 2000; Risaliti et al. 2002; 
Deluit \& Courvoisier 2003; Malizia et al. 2003), 
it must be kept in mind that an additional soft \xray\ component (perhaps 
related to the AGN host galaxy, to a circumnuclear starburst or to a 
scattering component; see, e.g., Vignali et al. 2001) could be present.
However, the redshift range of the present sample partially reduces the 
contribution of this soft component to the measured \hbox{0.5--2~keV} fluxes. 
For the sources where both the PSPC and HRI measurements are available, 
we compared the \xray\ fluxes (upper limits). These are usually similar 
(but see Table~1 for the case of source \#59), 
therefore we decided to use the PSPC measurements since they are available for 
the majority of the sources in the present sample. 
For the \xray\ detected source \#148, a variability of a factor of 
$\approx$~2 over time-scales of $\approx$~5~yr seems to be present. 
This is not surprising since \xray\ variability over time-scales of several 
years are often observed in AGN (Almaini et al. 2000; Markowitz, 
Edelson \& Vaughan 2003); furthermore, this source is also the most 
radio-loud AGN in Z03 sample (see Table~1), so enhanced variability, 
perhaps related to a jet component,  might be present. 
Also in this case the PSPC measurement is used. 

To match the Z03 sample with the RASS (Voges et al. 1999), we used the 
\rosat\ Web 
Browse\footnote{See http://www.xray.mpe.mpg.de/cgi-bin/rosat/src-browser.} 
assuming a matching radius of 40\arcsec\ and 
adopted a slightly more conservative detection threshold 
($L=8$, corresponding to the $\approx$~3.7$\sigma$ detection level) than 
that adopted for \rosat\ pointed observations. 
One candidate Type~II quasar is detected (at the $\approx$~6$\sigma$ level) 
in the RASS (source \#239 in Table~1). 
The detection of this source in the RASS was possible due to a longer than 
typical exposure ($\approx$~1100~s as compared to typical values of 
$\approx$~a few hundred seconds for most of the other Z03 sources). 
The \xray\ upper limits for the majority of the Z03 sources in the RASS 
are not sensitive enough for the purposes of this study. 

Recently, Zakamska et al. (2004) have searched for the counterparts of their 
Type~II AGN at radio/infrared wavelengths and in the \xray\ band 
(using the RASS only). 
They found six sources with a counterpart in the RASS 
\hbox{(at $0.31<z<0.61$)} using a matching radius of 1\arcmin\ 
(see their Table~4). 
Among these, source \#239 is in common with our list (see Table~1), 
while all of the remaining sources are below the detection threshold 
adopted in the present paper (three have $\simlt3.0\sigma$ detections, one 
is detected at the 3.4$\sigma$ level, and one at the 3.6$\sigma$ level). 
Two of these five matched sources have likely radio counterparts, 
thus increasing the reliability of their \xray\ detection in the shallow 
RASS fields. However, the visual inspection of the individual RASS images 
provides evidence for weak \xray\ emission only in the 3.4$\sigma$ RASS source 
(having also a radio counterpart). 
Including the additional five lower significance 
\hbox{($\approx2.1\sigma-3.6\sigma$)} RASS sources from Zakamska et al. (2004) 
would provide eight \xray\ detections of Type~II quasar candidates, 
although in the following we will refer only to the three higher significance \xray\ 
sources matching our more conservative selection criteria. 
As pointed out by Zakamska et al. (2004), it is interesting to note 
that the fraction of SDSS Type~I AGN (in the redshift range $0.3<z<0.8$) 
having RASS counterpart is significantly higher ($\approx$~15-20 per cent) 
than that of Z03 SDSS Type~II AGN. 
This can be explained assuming the presence of \xray\ obscuration 
($\approx2\times10^{22}$~cm$^{-2}$, Zakamska et al. 2004); a similar 
result is clearly found in $\S4$ of the present paper on a larger sample 
of Type~II quasar candidates with deeper \xray\ observations 
and also via direct \xray\ spectral analysis (for source \#204, see $\S$4.1).

\subsection{Chandra observations}
Only one source in the Z03 sample [\#130 (OBS\_ID=532);
also reported as HRI upper limit; see Table~1] 
lies within an archival \chandra\ observation (on-axis 
exposure time of $\approx$~8.1~ks) but is not detected. 
Unfortunately, its position, very close to the edge of 
ACIS-I field-of-view, prevents a reliable determination of the source 
\xray\ flux upper limit.

\subsection{XMM-Newton observations}
Two sources of the Z03 sample fall in \xmm\ archival observations. 
One is a clear \xray\ detection (source \#204, also detected by 
\rosat\ PSPC; see Table~1); the spectral analysis of this source using EPIC 
pn and MOS data is presented in $\S$4.1. 
The \xmm\ data were reduced with the version 5.4.1 of the 
Science Analysis Software ({\sc sas}) 
following the procedure described in Vignali et 
al. (2004). The data were filtered to avoid background flares 
which affect large intervals ($\approx$~60--70 per cent) of our observation. 
The tasks {\sc epproc} and {\sc emproc} were used to 
generate valid photon list; patterns 0--4 and 0--12 were used for pn and MOS, 
respectively. 
Source counts were extracted in the \hbox{0.3--10~keV} band 
from circular regions of radius 30\arcsec\ (pn) and 40\arcsec\ (MOS), 
while background counts were extracted from off-source regions of 
radius 90\arcsec\ (pn) and 80\arcsec\ (MOS). 
Response functions for spectral fitting were generated using the tasks 
{\sc rmfgen} and {\sc arfgen}. 

The other source [object \#59 (OBS\_ID=0111200101), also reported 
as PSPC upper limit; see Table~1] is not detected; 
its position, close to the edge of EPIC-pn 
field-of-view, does not allow us to derive a reliable \xray\ flux upper 
limit, despite the long exposure of the observation 
($\approx$~37~ks).

\subsection{Summary of X-ray observations}
Overall, three sources of the Z03 sample have \xray\ detections 
with \rosat\ (sources \#148, \#204, and \#239) assuming the 
matching criteria and detection thresholds reported in $\S$3.1. 
Source \#204, serendipitously found and detected also 
by \xmm, has enough counts for a moderate-quality spectral analysis; 
the most relevant results are presented in $\S$4.1.

\section{Constraints on the X-ray properties of Type~II quasar candidates}
The distribution of the \oiii\ luminosities (\loiii) vs. redshift for the 
Z03 sources is shown in Fig.~1. 
%
\begin{figure}
\includegraphics[angle=0,width=85mm]{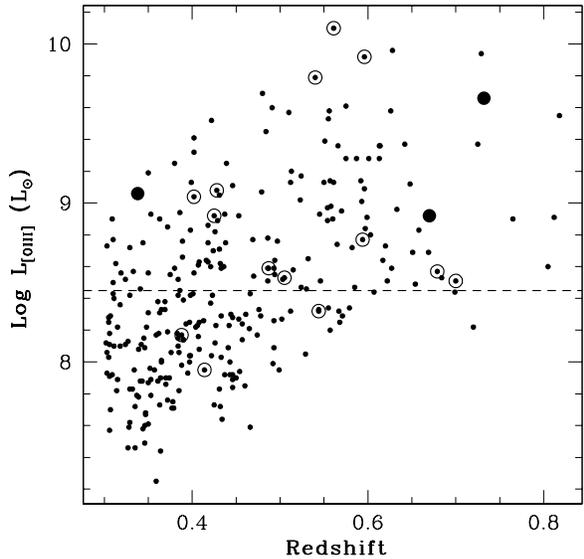}
\caption{
Logarithm of \loiii\ luminosities vs. redshifts for all of the sources in 
Z03 catalog (small filled circles). 
The large open circles indicate the objects with \xray\ upper limits, 
while the three \xray\ detections are shown as large filled circles. 
Objects above the dashed line have \oiii\ line luminosities in the range 
\hbox{$\approx$~3$\times10^{8}-10^{10}$~$\lsun$} 
which are comparable to those of luminous ($-27<M_{\rm B}<-23$) 
quasars; see Z03 for details.}
\label{fig1}
\end{figure}
%
Large open circles indicate the sources having \xray\ upper limits, while 
the three large filled circles indicate the objects with \xray\ detections. 
Most of the sources with \xray\ information 
have \oiii\ line luminosities in the range 
\hbox{3$\times10^{8}-10^{10}$~$\lsun$} 
(region above the dashed line in Fig.~1) 
which are comparable to those of luminous 
\hbox{($-27<M_{\rm B}<-23$)} quasars (Z03). 
In other words, {\em most of our \xray\ observed objects 
are Type~II quasar candidates} (see Footnote~2). 

To derive basic constraints on the \xray\ emission of Type~II 
quasar candidates, we have used the correlation between the \oiii\ flux 
and the \hbox{2--10~keV} flux found for Seyfert~II galaxies 
(Mulchaey et al. 1994), adopting the parameterization reported in $\S$3.2 
of Collinge \& Brandt (2000). 
Using this approach, we have computed the expected \hbox{2--10~keV} flux 
for each source and, using the 1$\sigma$ scatter in the correlation 
of Mulchaey et al. (1994), also the corresponding error range. 
It must be noted that no correction has been applied to the [OIII] line flux 
for the absorption due the narrow-line region itself 
[see Maiolino et al. (1998) and Bassani et al. (1999) for details]. 
This effect
is usually corrected for by using the Balmer decrement, but 
unfortunately the spectral coverage of the H$\alpha$ line is not available for 
most of the objects in the present sample.
This approach leads to a conservative determination in whether 
or not a source is absorbed in the \xray\ band. 

The predicted \hbox{2--10~keV} fluxes (with their 1$\sigma$ uncertainties) 
have been converted into soft \hbox{(0.5--2~keV)} \xray\ fluxes assuming a 
power law with photon index $\Gamma=2$, which seems to be a relatively good 
parameterization for the intrinsic \xray\ continuum of both Type~I and 
Type~II AGN (e.g., Nandra \& Pounds 1994; 
Deluit \& Courvoisier 2003).\footnote{The assumption of $\Gamma=1.9$ 
instead of 2.0 provides lower \hbox{0.5--2~keV} fluxes 
(by $\approx$~14 per cent).} 

The soft \xray\ fluxes derived following the method described above 
have then been compared with those observed 
(or the 3$\sigma$ upper limits for the \xray\ non-detections; see Table~1). 
This has allowed us to estimate on a source-by-source basis the amount 
of \xray\ absorption required to match the derived soft \xray\ flux 
with the observed one. 
Since the derived soft \xray\ flux has an associated scatter (due to that of 
the Mulchaey et al. 1994 correlation), for each source we determine an 
``allowed'' range of column densities ($N_{\rm H,z}-$ and $N_{\rm H,z}+$ in 
Table~2). 
For sake of completeness, the column density range associated with each 
source is shown in Fig.~2. 
%
\begin{figure}
\includegraphics[angle=0,width=85mm]{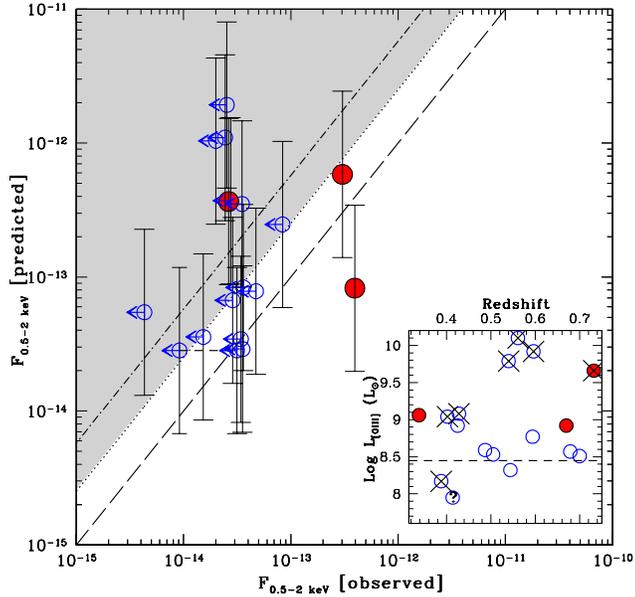}
\caption{
Predicted vs. observed soft \xray\ flux for Z03 objects with 
archival \xray\ observations. The filled circles indicate the 
three  \xray\ detections. 
Note the large \xray\ flux for the right-most object 
(source \#148) of the plot; 
this is also the most radio-loud source in the Z03 sample. 
The dashed diagonal line shows the 1:1 correlation, while the 
dotted (dot-dashed) line indicates the correlation 
expected for a source at $z=0.5$ if the column 
density is $1\times10^{22}$~cm$^{-2}$ ($3\times10^{22}$~cm$^{-2}$).
The horizontal dashed line connects the observed PSPC and HRI upper limits 
for source \#59. 
The grey region shows the locus of likely absorbed 
($N_{\rm H}>10^{22}$~cm$^{-2}$) sources. The column densities associated 
with our objects, given the method used for their derivation, 
should be considered conservative (see $\S$4 for details). 
In the inset, the \loiii\ vs. redshift (as Fig.~1) 
is plotted for the objects with \xray\ observations. 
The crosses indicate the likely absorbed objects 
(see Table~2); the ``?'' indicates source \#59, where absorption is probably 
present only in the HRI observation (see text and Table~2).}
\label{fig2}
\end{figure}
%
\begin{table}
\begin{minipage}{85mm}
\caption{Constraints on the \xray\ column densities derived 
from \rosat\ data.}
\begin{tabular}{@{}lcccc@{~~}c@{~~}c@{}}
\hline
Src & \multicolumn{3}{c}{Column Density$^{a}$} & X-ray$^{b}$ & 
Likely$^{c}$ & QSO~II$^{d}$\\
ID \# & \multicolumn{3}{c}{} & Det. & Abs. & \\
\cline{2-4} \\
 & \multicolumn{1}{c}{$N_{\rm H,z}$} & \multicolumn{1}{c}{$N_{\rm H,z}-$} & 
\multicolumn{1}{c}{$N_{\rm H,z}+$} & & & \\
\hline
{\phn}34  &  1.2$\times10^{23}$  &  6.0$\times10^{22}$  &     2.0$\times10^{23}$   &            &  $\surd$  &  $\ast$  \\ 
{\phn}55  &  4.6$\times10^{22}$  &  1.1$\times10^{22}$  &     9.6$\times10^{22}$   &            &  $\surd$  &          \\ 
{\phn}59  &    0                 &    0                 &     1.5$\times10^{22}$   &            &           &          \\ %
{\phn}59$^{e}$  
          &  1.2$\times10^{22}$  &  7.0$\times10^{21}$  &     4.9$\times10^{22}$   &            &  $\surd$  &          \\ 
{\phn}68  &    0                 &    0                 &     2.0$\times10^{22}$   &            &           &          \\
{\phn}70  &  8.0$\times10^{21}$  &    0                 &     4.7$\times10^{22}$   &            &           &          \\
     130  &  1.5$\times10^{23}$  &  8.0$\times10^{22}$  &     2.3$\times10^{23}$   &            &  $\surd$  &  $\ast$  \\ 
     148  &    0                 &    0                 &       0                  & $\ddagger$ &           &          \\ 
     152  &  5.0$\times10^{22}$  &  1.3$\times10^{22}$  &     1.0$\times10^{23}$   &            &  $\surd$  &  $\ast$  \\ 
     174  &    0                 &    0                 &     2.3$\times10^{22}$   &            &           &          \\
     188  &  4.3$\times10^{22}$  &  8.5$\times10^{21}$  &     9.5$\times10^{22}$   &            &  $\surd$  &  $\ast$  \\ 
     204  &  8.7$\times10^{22}$  &  2.3$\times10^{22}$  &     1.8$\times10^{23}$   & $\ddagger$ &  $\surd$  &  $\ast$  \\ 
     208  &  5.0$\times10^{21}$  &    0                 &     4.1$\times10^{22}$   &            &           &          \\
     209  &  9.0$\times10^{21}$  &    0                 &     4.6$\times10^{22}$   &            &           &          \\
     212  &  1.7$\times10^{22}$  &    0                 &     4.9$\times10^{22}$   &            &           &          \\ 
     239  &  4.7$\times10^{21}$  &    0                 &     3.0$\times10^{23}$   & $\ddagger$ &           &          \\
     256  &  1.3$\times10^{23}$  &  6.0$\times10^{22}$  &     2.1$\times10^{23}$   &            &  $\surd$  &  $\ast$  \\ 
     258  &  1.2$\times10^{22}$  &    0                 &     6.2$\times10^{22}$   &            &           &          \\
%
\hline
\end{tabular}
The column densities reported here (in cm$^{-2}$) should be considered as 
lower limits for all those objects which have not been detected by \rosat. 
For the three \xray\ detected sources, the column density range 
($N_{\rm H,z}-$, $N_{\rm H,z}+$) is broadly consistent with the absorption 
derived by the hardness-ratio analysis. \\
$^{a}$ The three values of column densities have been obtained 
using the best-fitting value and the corresponding scatter in the 
Mulchaey et al. (1994) correlation between the \oiii\ flux and the 
\hbox{2--10~keV} flux found for Seyfert~II galaxies; 
this correlation has been applied to our data 
and used to estimate the expected \hbox{0.5--2~keV} flux range 
for each object. Then the estimated \hbox{0.5--2~keV} fluxes have been 
compared with the observed \hbox{0.5--2~keV} fluxes 
(or 3$\sigma$ upper limits) to derive 
$N_{\rm H,z}$, $N_{\rm H,z}-$, and $N_{\rm H,z}+$ on a source-by-source basis. 
Given the procedures adopted to derive the column densities (see $\S$4 for 
details), $N_{\rm H,z}$ should be considered a conservative estimate. 
$^{b}$ $\ddagger$ means that the object is detected by \rosat; see Table~1. 
$^{c}$ $\surd$ marks the sources which are likely absorbed 
(i.e., $N_{\rm H,z}-~>0$); see text for details. 
$^{d}$ $\ast$ indicates the possibly absorbed sources with 
\xray\ luminosities $\gtrsim3\times10^{44}$~\lum (see Table~1 and 
Footnote~2).  $^{e}$ For this source, HRI data indicate the presence of 
absorption. The time interval between the PSPC and HRI observations 
is $\approx$~2~yr. 
\end{minipage}
\end{table}

%
This procedure is based on 
{\em the simplest assumption made possible by the data 
at our disposal, i.e., these Type~II quasar candidates have the same 
underlying average \xray\ continuum of local Seyfert~I galaxies and quasars, 
the only difference being the amount of \xray\ absorption, which has 
been assumed to be intrinsic to the source.} 
Clearly, the uncertainties associated with the predicted soft \xray\ fluxes 
(and thus with the derived column densities) are large. 
However, within our assumptions, 
these column densities (see Table~2) should be considered 
as lower limits, since all but three of our sources are not detected 
in the soft \xray\ band. 

At least 47 per cent (8/17) of the sources with \xray\ information 
shows indications of \xray\ absorption (see Fig.~2 and Table~2), 
with column densities typically 
$\simgt10^{22}$~cm$^{-2}$ (dotted line in Fig.~1 for a source at $z=0.5$). 
A column density of \hbox{$\approx$~1--2$\times10^{22}$~cm$^{-2}$}, 
close to the $N_{\rm H,z}-$ value, has been 
obtained also for source \#204, whose \xray\ properties have been 
derived via direct \xray\ spectral analysis using \xmm\ data 
(see $\S$4.1 and Fig.~3).

We also tried to perform a stacking analysis combining all of the \xray\ 
undetected sources. Although an $\approx$~4$\sigma$ detection has 
been obtained, the total number of \xray\ source counts ($\approx$~34) 
did not allow us to derive an average column density 
through hardness-ratio analysis. 
However, we compared the average flux derived from the stacking analysis 
(weighted by the exposure time of each observation with respect to the total) 
with that expected on the basis of the correlation of Mulchaey et al. (1994). 
An average column density of \hbox{1.4--2.7$\times10^{23}$~cm$^{2}$} has been 
derived, although we note that this value suffers from significant 
uncertainties. More reliable constraints could be obtained using \chandra, 
given the lower background and sharper PSF. 

A caveat to the plain use of the method described above 
is related to the fact that the observations in the optical 
and \xray\ bands are not simultaneous. 
Intrinsic source variability as well as variability 
in the absorbing medium over the time-scales probed by the 
\xray\ observations might be present. 
However, our approach, although limited by the paucity of \xray\ detections, 
is simple and designed to place basic constraints on the 
overall \xray\ properties of optically selected, Type~II quasar candidates. 
%
\begin{figure*}
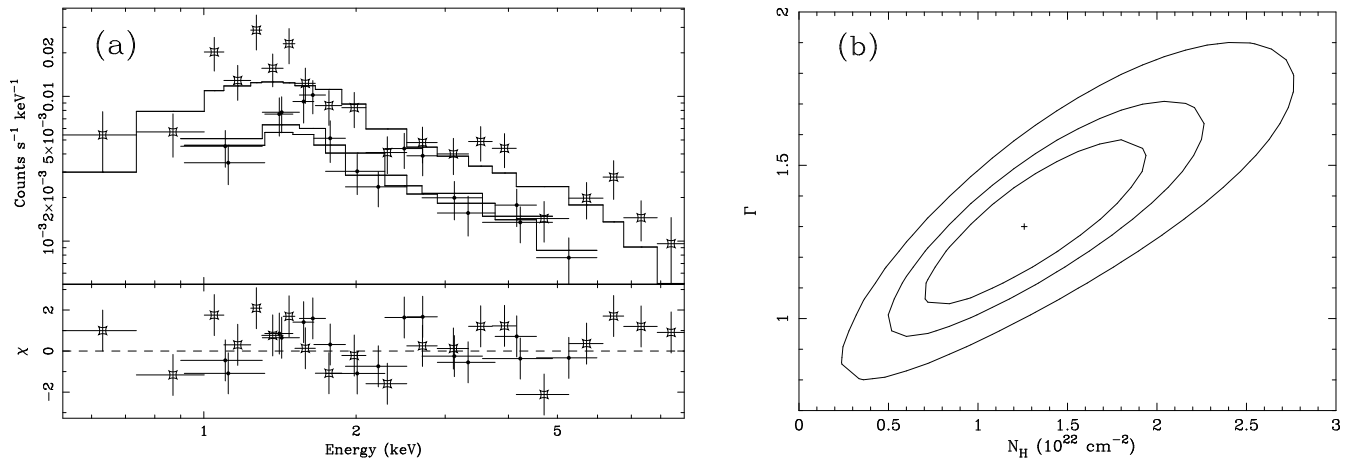

\includegraphics[width=60mm,angle=-90]{ME601rv2.fig3a.ps}
\hfill
\includegraphics[width=60mm,angle=-90]{ME601rv2.fig3b.ps}
\caption{
\xmm\ data of source \#204: (a) pn (larger symbols) $+$MOS1$+$MOS2 (small 
filled circles) spectral data (top panel) and data/model residuals 
(bottom panel, in units of $\sigma$); 
(b) 68, 90, and 99 per cent confidence contours for the photon index vs. 
intrinsic column density; see $\S$4.1 for details.} 
\label{fig3}
\end{figure*}
%

\subsection{X-ray spectral analysis of source \#204: XMM-Newton results}
Source \#204 has $\approx510$ counts in the pn$+$MOS1$+$MOS2 data. 
These counts were grouped into spectra such that each spectral bin 
contained at least 15 counts to allow $\chi^{2}$ fitting. 
\xray\ spectra were fitted using the {\sc xspec} package (version 11.2; 
Arnaud 1996). The quoted errors on the derived model parameters correspond to 
the 90 per cent confidence level for one interesting parameter 
(i.e., \hbox{$\Delta\chi^{2}=2.71$}; Avni 1976). 
All spectral fits include absorption due to the line-of-sight Galactic column 
density (from Dickey \& Lockman 1990; see Table~1). 
Separate spectral fitting indicates that pn and MOS spectral results are 
consistent to within $\approx$~10 per cent; this is similar to the value 
quoted by Kirsch (2003; $\approx$~13 per cent). Therefore in the following 
the pn and MOS data will be fitted together with the same model 
leaving the relative normalisations free to vary. 

A single power-law model is clearly not an adequate fit to \xmm\ data; 
the quality of the fit is poor \hbox{($\chi^{2}$/d.o.f.=61.2/32)} and the 
photon index is very flat \hbox{($\Gamma=0.74^{+0.12}_{-0.13}$)}, 
suggesting the presence of absorption. 
If an absorber at the source redshift \hbox{($z=0.732$)} is 
added to the previous model, 
the fit improves significantly ($\chi^{2}$/d.o.f.=45.4/31), although 
some residuals are still present at energies around 1.5~keV. 
These can be partially accounted for by the addition of a thermal component 
({\sc mekal} model in {\sc xspec}), perhaps related to the AGN host galaxy or 
to a circumnuclear starburst, but the spectral parameters of this component 
are poorly constrained, and therefore it will not be considered 
in the following.  
The best-fitting spectrum [shown in Fig.~3, panel (a)] 
is parameterized by a photon index 
\hbox{$\Gamma=1.30^{+0.16}_{-0.24}$} and a column density 
\hbox{$N_{\rm H}=1.26^{+0.75}_{-0.51}\times10^{22}$~cm$^{-2}$}. 
The 68, 90, and 99 per cent confidence contours for the photon index 
vs. column density are shown in Fig.~3 [panel (b)]. 
If the photon index is forced to be \hbox{$\Gamma=1.8-2.0$}, 
more representative of the typical AGN intrinsic \xray\ emission, 
the column density becomes \hbox{2.2--3.2$\times10^{22}$~cm$^{-2}$}, 
more consistent with the $N_{\rm H}$ range derived from \rosat\ data 
and reported in Table~2. 
We note, however, that source variability may have occurred between the 
\rosat\ and \xmm\ observations (over a time-scale of $\approx$~10~yr). 
The intrinsic \hbox{2--10~keV} luminosity \hbox{(10$^{44.6-44.7}$~\lum)}, 
the column density \hbox{($\approx$~1--2$\times10^{22}$~cm$^{-2}$)}, 
and the narrow-line optical spectrum (Z03) 
are fully consistent with source \#204 being a genuine Type~II quasar.

\section{Discussion}
Though not complete, the catalog of Z03 provides 
an optically well-defined sample of candidate Type~II quasars which can be 
used as a starting point for detailed analyses of the broad-band properties 
of optically luminous Type~II quasars and for the study of the contribution 
of this population to the hard XRB (e.g., Gilli et al. 2001). 
In particular, Type~II quasars might be crucial to study the 
dependence of absorption upon \xray\ luminosity (Ueda et al. 2003; 
La Franca et al., in preparation). 
Since the sources of the present sample are mostly optically luminous, 
this study can be considered complementary to that of \xray\ selected Type~II 
quasar candidates, whose optical counterparts are typically faint and often 
hamper an accurate spectroscopic identification. 
Clearly, the results presented in this paper suffer from significant 
uncertainties, mostly related to the scatter in the correlation 
between the \oiii\ line intensity and hard \xray\ flux (Mulchaey et al. 1994), 
the applicability of this method to the higher luminosity ``cousins'' of 
local Seyfert~II galaxies (from which the correlation is obtained), and 
the assumption that all the AGN spectra have on average 
the same $\Gamma\approx2$ underlying \xray\ continuum. 
The first of these three major sources of uncertainty has been taken into 
account by comparing the observed soft \xray\ fluxes and upper limits with 
those obtained using the correlation above. 
The uncertainties related to the scatter in the correlation 
(vertical error bars in Fig.~2) 
affect the derivation of the column densities 
(see the values $N_{\rm H,z}-$ and $N_{\rm H,z}+$ in Table~2); 
it must be kept in mind, however, that the predicted soft \xray\ fluxes have 
been compared mostly with the observed \xray\ flux upper limits, 
therefore the column densities should be treated as lower limits. 
Furthermore, intrinsically flatter \xray\ slopes would produce lower 
column densities (by $\approx$~20 per cent, if $\Gamma=1.6$ instead of 2.0 
is adopted), while the presence of an additional soft \xray\ component 
(e.g., thermal emission from the host galaxy, scattering, etc.) 
would produce the opposite effect. 
Given the lack of spectral constraints provided by direct \xray\ fitting 
(except for source \#204), this approach, although a bit simplistic, 
allowed us to derive basic \xray\ constraints for 17 Type~II AGN, some of 
which being quasar candidates: \\
{\em (i)} At least 47 per cent of the present sample shows indications 
of \xray\ absorption, with column densities $\gtrsim10^{22}$~cm$^{-2}$. \\
{\em (ii)} 
The presence of genuine Type~II quasars among the objects presented in this 
paper finds further support from the only object (source \#204) 
with \xmm\ moderate-quality spectral data ($\S$4.1). 
The intrinsic \hbox{2--10~keV} luminosity \hbox{($\approx$~10$^{44.6}$~\lum)}, 
the column density \hbox{($\approx$~1--2$\times10^{22}$~cm$^{-2}$)}, 
and the narrow-line optical spectrum of this source are consistent with 
the Type~II quasar definition. 
Moreover, its \xray\ luminosity is comparable with those of some 
higher redshift \xray\ selected Type~II quasars 
(e.g., Stern et al. 2002; Mainieri et al. 2002; Brusa et al. 2003).

Our results find further support from those obtained recently by 
Zakamska et al. (2004) on the basis of RASS data only. 

The large-area surveyed by the SDSS, coupled with the relatively bright flux 
limit, the spectroscopic target selection and wavelength coverage, is such 
to possibly provide large numbers of optically selected Type~II quasar 
candidates to be followed-up at other wavelengths with, e.g., \spitzer, 
\scuba\, and current \xray\ telescopes such as \chandra\ and \xmm. 
Therefore, in the following years it will become possible to properly assess 
the broad-band properties of optically selected Type~II quasars 
(in particular, their spectral energy distribution) and estimate 
their contribution to the hard XRB.

\section*{Acknowledgments}
CV and AC acknowledge partial support by the Italian Space agency 
under the contract ASI I/R/057/02. DMA is supported by the Royal Society. 
The authors would like to thank L. Angeretti for help with the plots, 
M. Mignoli and L. Pozzetti for useful discussions, and G. Zamorani for 
a careful reading of the manuscript. 
CV gratefully remembers Rosa B.

\end{document}